\documentstyle[12pt,epsfig]{article}

\begin{document}

\begin{flushright}
CERN-TH/99-314\\
gr-qc/0001100 \\
%$~$ \\
%October 1999
\end{flushright}

\begin{centering}
\bigskip
{\Large \bf A testable description of space-time foam as a\\
fundamental stochastic gravity-wave background}\\
\bigskip
\bigskip
\bigskip
{\bf Giovanni AMELINO-CAMELIA\footnote{{\it Marie Curie
Fellow} of the European Union
(address from February 2000: Dipartimento di Fisica,
Universit\'a di Roma ``La Sapienza'',
Piazzale Moro 2, Roma, Italy)}}\\
\bigskip
Theory Division, CERN, CH-1211, Geneva, Switzerland \\ 
\end{centering}
\vspace{1cm}
\begin{center}
{\bf ABSTRACT}
\end{center}

{\leftskip=0.6in \rightskip=0.6in

I develop a phenomenological approach
to the description of the noise levels that
the space-time foam of quantum gravity could induce
in modern gravity-wave detectors.
Various possibilities are considered, including
white noise and random-walk noise.
In particular, I find that
the sensitivity level expected for the planned 
LIGO and VIRGO interferometers and for the
next upgrade of the NAUTILUS resonant-bar detector
corresponds to a white-noise level which can be 
naturally associated with the Planck length.

}%%%%%

\baselineskip = 12pt
     
\vspace{1cm}

One of the most natural expectations for quantum gravity,
as the theory describing the interplay
between gravity and quantum mechanics,
is that space-time, when resolved at very short distances, would appear
to be ``foamy'' in the sense of Refs.~\cite{wheely,hawkfoam}.
The fact that there is still a rather wide collection of intuitions for
this fascinating new picture of space-time (see, {\it e.g.}, the recent
Refs.~\cite{fotleefoam,ambjfoam}, which also provide a good
starting point for a literature search backward in time)
is due to the technical and conceptual difficulties
encountered in the development of theoretical approaches
to the quantum-gravity problem.
Most quantum-gravity theories have not yet
passed even the most basic tests of consistency.
The two approaches that have survived at least a
few non-trivial consistency tests,
the one based on ``critical superstrings''~\cite{string1,string2}
and the one based on ``canonical/loop quantum
gravity''~\cite{cqglee,cqgab,cqgcar},
do not have any direct confirmation from experimental data
and even theoretical studies of the nature
of their physical implications 
are only at a preliminary stage.

While waiting for the emergence of a ``full-grown''
quantum gravity, possibly through the maturation
of one of the mentioned approaches,
it is becoming increasingly clear that,
by exploiting recent
progress in experimental technologies and ideas,
we can follow an alternative path~\cite{elmn,grbgac,gacgwi} 
toward the exploration of space-time foam.
It has been shown~\cite{elmn,grbgac,gacgwi} 
that certain types of experiments
have become so refined that their sensitivities 
can be naturally expressed as proportional 
to the Planck length $L_p \sim 1.6 \cdot 10^{-35}m$ 
(whose smallness we expect to 
penalize all quantum-gravity effects).
In order to profit from these new experimental possibilities
one can set up phenomenological models providing
estimates (typically depending
on a few unknown parameters encoding our ignorance of
the correct quantum gravity) of candidate quantum-gravity
effects. The hope is that these phenomenological
estimates may guide experimentalists
toward the discovery of some quantum-gravity phenomena,
which in turn would
provide much needed hints for the rigorous mathematical
work searching for the correct quantum-gravity formalism.
[More detailed considerations on the impact
that this phenomenological approach might have
on the development of quantum gravity can be found in
Refs.~\cite{cqglee,cqgab,cqgcar,ahlunature,divulg}.]

In one of these phenomenological proposals
I observed~\cite{gacgwi} that the quantum fluctuations 
affecting distances in conventional pictures of space-time foam
would manifest themselves in the operation of
modern gravity-wave detectors in a way that mimics a
stochastic gravity-wave background.
Just like a stochastic gravity-wave background 
these quantum-gravity effects would
induce stochastic fluctuations
in the magnitude of distances, and just
like a stochastic gravity-wave background 
these quantum-gravity effects would
be felt in a sensitive gravity-wave detector
as an additional source of noise.
I also observed that, as done for ordinary stochastic 
gravity-wave backgrounds, the power spectrum of the 
strain noise~\cite{saulson} that
would be induced in gravity-wave detectors
is the most convenient way
to characterize models of foam-induced
distance fluctuations. This predicted
strain noise power spectrum
can be compared to the
strain noise power spectrum actually found in a
given detector, thereby obtaining
bounds on the parameters of phenomenological
descriptions of the foam-induced
distance fluctuations.

In the present Letter I use these observations
as motivation for a phenomenological approach to the study
of space-time foam in which 
some properties of foam are modeled 
as a fundamental/intrinsic level of 
strain noise power spectrum. It may seem hard to 
develop a phenomenology directly at the level of the foam,
without an underlying theory of quantum gravity, but
I shall show that
the assumption that an appropriate characterization be given by a 
strain noise power spectrum (which is a function of a 
single variable)
together with the assumption
that this fundamental level of strain noise
should be a universal property of physics 
(in a sense that will become clearer below) 
provide rather strong constraints for the construction
of candidate power spectra.

Let me start by discussing the possibility that
this foam-induced
strain noise power spectrum
be due to underlying 
quantum-gravity space-time fluctuations that are 
of random-walk type.
This is a rather simple hypothesis which also
fits well the intuition emerging from certain
approaches to the more formal analysis of space-time foam
(see, {\it e.g.}, Refs.~\cite{fotleefoam,ambjfoam}
and references therein).
From this simple hypothesis it already follows that
the functional form of the
strain noise power spectrum is\footnote{I introduce 
the index $h$ on $\rho_h$
only to emphasize that this is the strain noise 
power spectrum ($h$ is conventionally used to denote the strain),
rather than the distance noise 
power spectrum. The distance noise 
power spectrum contains the same basic information
and for most applications can be obtained multiplying $\rho_h$ 
by the square of an appropriate length scale in the detector
(which for interferometers is given by the length of the arms).
Also note that power spectra are most commonly
denoted by the symbol ``$S$'', but
I prefer to use the symbol ``$\rho$'' because the symbol ``$S$''
is becoming somewhat ambiguous 
as a result of the fact that some authors have used ``$S$'' for 
the ``amplitude spectral density'' which is the square root
of the power spectrum.}
\begin{equation}
\left[ \rho_h(f) \right]_{random~walk} \sim {1 \over f^2}  ~.
\label{rwgeneral}
\end{equation}
There is in fact a general correspondence
(see, {\it e.g.}, Refs.~\cite{rwold,polonpap})
between processes of random-walk type
and power spectra with $f^{-2}$ frequency dependence.
In Refs.~\cite{gacgwi,polonpap} I observed that
one could find several plausible (though in a certain
sense ``optimistic''~\cite{polonpap}) 
quantum-gravity arguments
in support of the candidate foam-induced strain noise power spectrum
\begin{equation}
\rho_h(f) = {L_p c \over L^2} {1 \over f^2}  ~,
\label{rwgwi}
\end{equation}
where $c \simeq 3 \cdot 10^8 m/s$ is the speed-of-light 
constant and $L$ is 
an appropriate length scale characterizing the detector,
which for interferometers is given by the length of the arms,
but the sensitivity achieved with the {\it Caltech 40-meter
interferometer}~\cite{ligoprototype}
rules out~\cite{gacgwi,polonpap}
the possibility (\ref{rwgwi}).
The fact that the candidate (\ref{rwgwi}) could be ruled out
in spite of the minuteness of the Planck length
was used in Refs.~\cite{gacgwi,polonpap} 
to argue that modern gravity-wave detectors have reached
a level of sophistication such that it is no longer 
implausible that they might detect some 
(stochastic-gravity-wave-like)
quantum properties of space-time.
Similar conclusions have been drawn more recently in
Refs.~\cite{ahlunature,nggwi,adriangwi}.

While these earlier papers had clarified that some
exploration of quantum properties of space-time
is possible, the phenomenological approach
advocated in the present Letter should contribute
to define in a semiquantitative fashion
how far this exploration can go using planned gravity-wave
detectors.
As a warm-up exercise let us rederive 
(\ref{rwgwi}) from a direct phenomenological
analysis rather than from 
one sort or another of quantum-gravity arguments.
Let us in particular observe that 
by simply assuming that the underlying processes 
should be of random-walk type
(which, as mentioned, implies $\rho_h$ proportional to $f^{-2}$)
and that the relevant quantum-gravity effect
should be linear in the Planck length
(which is the most optimistic plausible possibility~\cite{polonpap}) 
the form of the power spectrum $\rho_h$ is completely
specified up to an overall coefficient with 
dimensions $m^{-1} \cdot s^{-1}$.
Having already assigned the dependence on $L_p$ and $f$,
it is clear that this overall coefficient must be constructed
out of the speed-of-light constant and some length scale
(a sort of cut-off scale) 
characterizing the relevant physical context.
In an interferometer a rather conservative estimate
for this coefficient is given indeed by $c/L^2$
(compare with (\ref{rwgwi})), since the length of the arms $L$
is the largest length scale in the physical context,
but even this most conservative
estimate turns out to be ruled out.
In a sense the phenomenological analysis of the random-walk noise 
scenario completely\footnote{Note however that,
while the coefficient $c/L^2$
is the smallest obtainable with a single one of the length scales
in an interferometer, 
it is of course possible to obtain smaller 
coefficients multiplying $c/L^2$ by small dimensionless ratios of other
length scales of the interferometer. 
An interferometer is complex enough 
(with its many length scales, such as the wavelength of the 
laser beam, the width of the laser beam, the size of the mirrors,...)
that such coefficients cannot be excluded; however, a prejudice 
of the present phenomenological analysis is that the 
noise induced by space-time foam should be a fundamental
property of {\it Nature} and that (as desirable for such 
fundamental entities) it should not have complicated
dependence on the specific physical context.}
rules out the possibility of an effect linear in the Planck length
because even the most conservative estimate of the coefficient
turns out to give too much noise with respect to the astonishing level 
of sensitivity achieved by modern gravity-wave detectors.
Of course, it remains possible that we have indeed foam-induced
strain noise of random-walk type, but that it be quadratically 
suppressed by the Planck length; I shall further
comment on this possibility in my closing remarks.

While random-walk noise fits well the intuition of some theoretical
approaches~\cite{fotleefoam,ambjfoam} to space-time foam, other
popular quantum-gravity ideas 
provide an intuition about space-time foam (see Ref.~\cite{garaythermal},
where an attempt was made at a formalization of this intuition,
and references therein) 
which might support the possibility of white noise 
since it establishes a certain level of analogy
between space-time foam and a thermal environment.
Moreover, one would anyway want to consider
the possibility of quantum-gravity-induced white noise
since white noise is the most common type of noise in {\it Nature},
especially for the low-frequency limit of power spectra
associated with processes whose characteristic 
frequencies are very high (and of course the characteristic 
frequencies of quantum-gravity processes are likely to be extremely
high since these processes occur on very short distances).
For the case of white noise the assumption that the effect be linear
brings about a remarkable simplicity in the description of the
power spectrum; in fact, by assuming that the noise be white
({\it i.e.} with frequency-independent power spectrum),
and that the power spectrum be proportional to the Planck length,
the form of $\rho_h$ is completely
specified up to an overall coefficient with 
dimensions $s \cdot m^{-1}$.
Of course, a compellingly simple 
choice of this overall coefficient is given 
by the inverse of the speed-of-light constant,
in which case one obtains 
\begin{equation}
 \rho_h(f) = {\rm constant} =  {L_p \over c} 
\sim 5 \cdot 10^{-44} Hz^{-1} ~.
\label{white}
\end{equation}
Notice that here, because
of the dimensionality of the overall coefficient,
the natural estimate does not
involve any length scale
characteristic of the physical context. This appears to
be a ``pleasant'' property. Quantum-gravity-induced 
stochastic-gravity-wave-like noise would be independent of the 
physical context, an intrinsic property of space-time.
It also allows us to compare the same estimate (\ref{white})
with both data obtained with interferometers and data
obtained with resonant-bar detectors~\cite{nautilus}
(in cases in which instead the estimate depends on length scales
characteristic of the detectors, 
one expects, of course, 
different independent estimates for interferometers and 
for resonant-bar detectors).
I also observe that,
while it is easy to verify that all 
presently-available data~\cite{ligoprototype,nautilus}
are comfortably consistent with the estimate (\ref{white}),
some of the detectors that will start operating soon
are expected to achieve sensitivity even beyond
the one required to test the estimate (\ref{white}).
Improvements in the NAUTILUS resonant-bar detector
are expected~\cite{nautilus,micgwb}
to reach sensitivity
at the level $7 \cdot 10^{-45} H\!z^{-1}$
within a few years.
The LIGO/VIRGO generation of 
interferometers~\cite{ligo1,ligo2,virgo1,virgo2} 
should achieve sensitivity just of the required 
order of $10^{-44} H\!z^{-1}$ within a year or two,
and a few years later both
the ``advanced phase''~\cite{micgwb,ligo2} 
of the LIGO/VIRGO interferometers and
the space interferometer LISA~\cite{lisa}
should improve the sensitivity by at least two or three
additional orders of magnitude.
It is amusing to notice that all these machines have been tuned 
to reach sensitivities in the neighborhood of $10^{-44} H\!z^{-1}$
because their primary objective is the discovery of the classical-physics
phenomenon of gravity waves, predicted by Einstein's general relativity,
and it just happens to be the case that
the relevant classical-physics studies have led to the conclusion that
a sensitivity somewhere between $10^{-42} H\!z^{-1}$ 
and  $10^{-46} H\!z^{-1}$
is needed for the discovery of classical gravity waves.
It is a remarkable numerical accident that the result of these
classical-physics studies pointed us toward a sensitivity level
which I am now observing to be also naturally associated with the
intrinsically quantum scale $L_p/c$. 
[It is perhaps worth noticing, incidentally, 
that, even setting aside
the intuition concerning space-time foam
advocated in the present Letter,
it is quite significant that no previous work
on modern interferometers had emphasized 
that $10^{-44} H\!z^{-1}$ is roughly of order $L_p/c$.
Clearly, the realm of quantum gravity 
is still quite distant~\cite{polonpap} 
from the set of ideas considered by most experimentalists.]

Having considered the case of random-walk ($f^{-2}$) noise
and white ($f^{0}$) noise, I shall next emphasize that 
it is also interesting to investigate
foam-induced noise scenarios somewhere between these two extremes,
scenarios with power spectrum going like $f^{-2 \beta}$
with $0 < \beta < 1$.
For example,
from a phenomenological model with  $\beta = 1$ (random-walk) noise
one can easily obtain a phenomenological model  
with $0 < \beta < 1$ by introducing some sort of ``pull-back
mechanism'' for space-time (some mechanism that tends to restore
space-time toward some original state if quantum fluctuations
have added up to a very large deviation).
Of particular significance
is the $\beta = 1/2$ case, with strain noise
power spectrum going like $f^{-1}$,
since from the studies reported in Refs.~\cite{jare}
one can infer~\cite{polonpap} that the contribution
of gravitons to quantum-gravity-induced noise
should have this type of behavior and should 
be proportional to the square of $L_p$.
While it is hard to imagine that this graviton contribution
could be detected in the foreseeable future
(dimensional analyses analogous to the ones presented above
lead to extremely small estimates of noise levels when applied
to a spectrum of $L_p^2/f$ type)
it is plausible that other (possibly nonperturbative)
aspects of quantum gravity might also lead to an $f^{-1}$ behavior.
However, it is easy to verify that the data reported in
Ref.~\cite{ligoprototype} already rule out the possibility
$\rho_h \sim L_p/(L f)$, which corresponds to $f^{-1}$ noise
proportional to (only one power of) the Planck length.

Another special case, among those with $0 < \beta < 1$,
is the one with $\beta = 5/6$
(strain noise power spectrum going like $f^{-5/3}$)
which appears to be preferred by certain arguments 
combining quantum measurement analysis and classical-gravity 
estimates~\cite{ng} as well as arguments concerning a 
scenario for a quantum-gravity measurement theory~\cite{polonpap}.
These arguments would lead specifically~\cite{gacgwi,polonpap} 
to the strain noise power spectrum $L_p^{4/3} c^{2/3}/(L^2 f^{5/3})$,
which is somewhat beyond the reach of the LIGO/VIRGO generation
of interferometers (for LIGO/VIRGO arms of order 3 or 4 $Km$
this would give $\rho_h \sim 10^{-51} Hz^{-1}$ at $f \sim 100 Hz$ 
while, even in their advanced phase, LIGO/VIRGO are not expected
to reach below $10^{-48} Hz^{-1}$)
and would also be beyond the reach of LISA\footnote{In a sense,
for what concerns the
estimate $\rho_h \sim L_p^{4/3} c^{2/3}/(L^2 f^{5/3})$,
LISA gains a factor of roughly $10^{9}$ with respect
to LIGO/VIRGO by going to lower frequencies but looses a factor of
roughly $10^{12}$ due to the much greater length $L$ of its arms.}.
Interestingly, however, LIGO/VIRGO could be 
sensitive~\cite{gacgwi,polonpap}
to $\rho_h \sim L_s^{4/3} c^{2/3}/(L^2 f^{5/3})$ 
with $L_s \sim L_p/20$,
which is obtained from the previous estimate by replacing 
the Planck length with a plausible value of the string length.

Before closing, it is worth going back to the case of 
foam-induced random-walk strain noise, just to notice that
with LISA we will start to perform meaningful
studies of the case in which 
the square of the Planck length sets the strength
of this noise scenario. 
The assumption that the underlying processes 
should be of random-walk type and that the relevant 
quantum-gravity effect should be proportional to the
square of the Planck length
implies that $\rho_h \sim L_p^2/f^2$ 
up to an overall coefficient with 
dimensions $m^{-2} \cdot s^{-1}$.
As mentioned above, a plausible scale to appear in these overall
coefficients is the length of the arms of the interferometer.
Other plausible length scales are the ones characterizing 
properties of the laser beam, like the laser wavelength $\lambda$
(typically $\lambda \sim 10^{-6}m$)
and the width of the beam $W$ (typically of a few $cm$).
This second type of ``cut-off scales'' would 
also fit the general expectation that truly non-perturbative
effects, such as the ones here considered, are likely
to manifest themselves at low energies through ratios 
of their characteristic scale ($L_p$) and a scale characterizing 
the size of the probes being used ({\it e.g.} the wavelength of
the photons).
While LISA will not be able to probe the case in which the 
overall coefficient is fixed by the length of the arms of 
the interferometer ($\rho_h \sim c L_p^2/(L^3 f^2)$ is safely
beyond the reach of LISA), significant sensitivity will be achieved
with respect to the case in which the 
overall coefficient is obtained from a combination of scales
characteristic of the laser beam. 
For example, LISA sensitivities will suffice to test random-walk 
scenarios with strength between $c L_p^2/\lambda^3$
and $c L_p^2/(\lambda W^2)$
({\it i.e.} power spectra
between $\rho_h \sim c L_p^2/(\lambda^3 f^2)$
and $\rho_h \sim c L_p^2/(\lambda W^2 f^2)$).

In summary, within the strictly phenomenological
approach advocated here the outlook of 
quantum-space-time studies by gravity-wave detectors
appears to be quite interesting.
As shown in the figure, the progress of this exploration will be 
extremely rapid over the next 10 or 15 years, which will take us
from the present interferometers (sophisticated machines but
still with arms of ``only'' 30 or 40 meters)
through LIGO and VIRGO (with arms of 4 and 3 $Km$ respectively)
all the way to the truly gigantic LISA interferometer
(with arms of $5 \cdot 10^6 Km$).
As emphasized here, we are getting ready to pass
some significant ``natural milestones'' of this exploration,
which are set by compellingly simple
combinations of fundamental constants.
Most notably the $\rho_h \sim L_p/c$ barrier
will soon be crossed by LIGO/VIRGO.
Also significant, especially in light of the fact that all 
other quantum-gravity 
experiments~\cite{elmn,grbgac,gacgwi,polonpap,schaef}
concern effects that are
only linearly penalized by the smallness of the Planck length,
is the fact that with LISA we will start to perform meaningful 
studies of noise levels suppressed by the square of the Planck 
length (for the case of random-walk noise).

Besides providing models of some properties of space-time
foam, the phenomenological approach here advocated 
could also be useful in bridging the gap between
experiments and theory concerning foam-induced noise.
In fact, clearly gravity-wave detectors are our most
sensitive probes of possible space-time fluctuations
and eventually quantum-gravity theories predicting fuzzy/foamy
space-times should provide to the experimentalists estimates
of the strain noise power spectrum.
The simple type of phenomenological models of space-time
foam here considered can also be used to describe
in terms of the characteristic Planck length the level of
sensitivity that various experiments are reaching 
with respect to certain types of noise.

As a final remark, let me observe that it
would be important for the development of the
phenomenological approach here advocated if 
it became clear how to apply ``energy constraints''
to the strain noise power spectra that the approach generates.
Had we been considering noise that could be attributed to gravitons
then standard energy constraints would apply, and these
can be quite restrictive (in particular,
by requiring not to overclose the {\it Universe} one 
would in that case 
find that white noise at the $L_p/c$ level 
could not extend above a few hundred $Hz$).
However, in this phenomenological approach one 
would like to maintain the analysis as model-independent
as possible, and actually, because of the nature of 
the considerations being made, 
in finding support for this type of phenomenology
one should be looking beyond gravitons,
considering foam-induced noise 
as a truly fundamental intrinsic property of space-time 
rooted in the fully non-perturbative structure of the 
(yet-to-be-discovered) theory.
For ``non-graviton'' 
gravitational degrees of freedom energy considerations 
are non-trivial. These subtleties
of quantum gravity with respect to issues related to
energy conservation have been elegantly reemphasized in 
the recent Ref.~\cite{thoodete}.
A very important tool for the phenomenology here advocated
would become available if we managed to find our way through
these delicate conceptual issues.

\baselineskip 12pt plus .5pt minus .5pt

\baselineskip = 12pt
     
\section*{Acknowledgements}
I am indebted to
Ram Brustein, Michele Maggiore and Gabriele Veneziano
for tutoring me on some aspects of the physics
of stochastic gravity-wave backgrounds.
Still on the theory side I am also grateful to several
colleagues who provided encouragement and stimulating
feed-back, particularly Dharam Ahluwalia, Abhay Ashtekar,
John Ellis, Nick Mavromatos, Jorge Pullin, Carlo Rovelli,
Subir Sarkar, Lee Smolin and John Stachel.
On the experiment side I would like to thank
J\'er\^ome Faist, Peter Fritschel, Luca Gammaitoni, 
Lorenzo Marrucci, Soumya Mohanty,
and Michele Punturo,
for conversations 
on various aspects of interferometry.

\vfil 
\eject

\newpage
\pagestyle{empty}

\begin{figure}[p]
\begin{center}
\epsfig{file=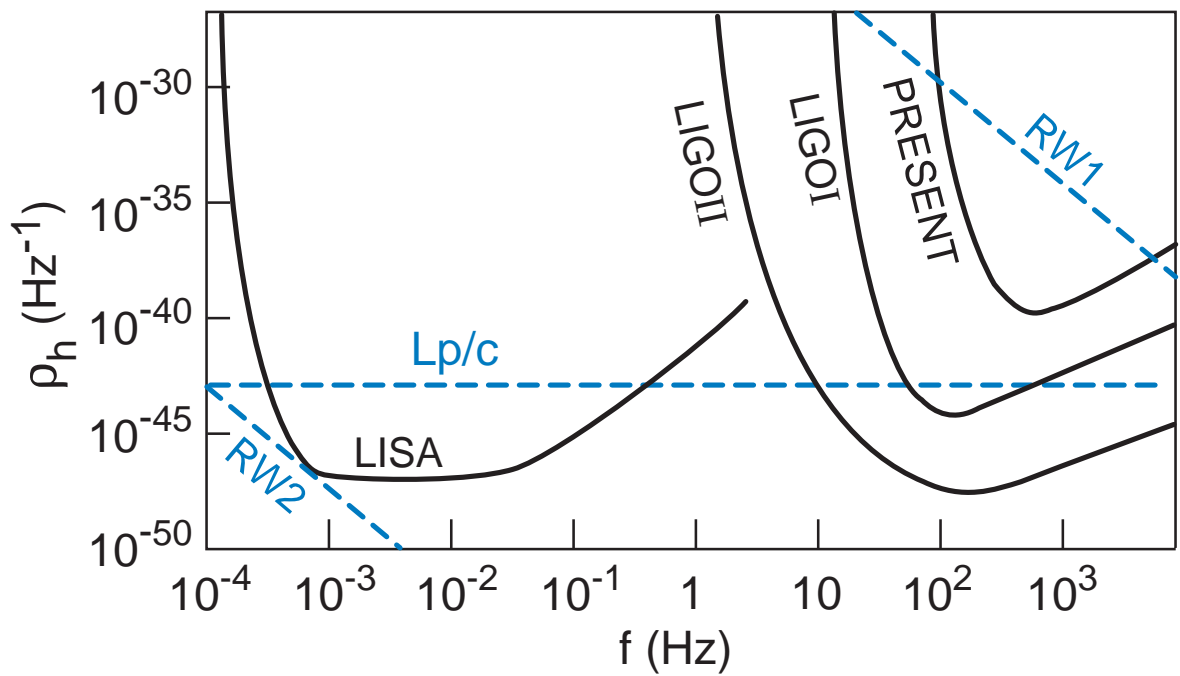,height=9truecm}\quad
\end{center}
\bigskip
\caption{A qualitative
(at best semi-quantitative) comparison
between the sensitivity of some interferometers 
and some of the phenomenological 
strain noise power spectra here considered.
The evolution from the level of sensitivity (``PRESENT'')
of interferometers already in operation,
to the first phase of the LIGO and VIRGO interferometers (``LIGOI''),
then to the second phase of LIGO and VIRGO (``LIGOII''),
and finally to LISA (``LISA'')
will take us through some significant phenomenological milestones
among candidate foam-induced noise levels.
The line ``RW1'' corresponds to the random-walk scenario 
(mentioned in the text)
with magnitude suppressed linearly by the Planck length,
and is clearly ruled out by ``PRESENT'' data.
The line ``$L_p/c$'' corresponds to the scenario with 
white noise at the $L_p/c$ level and it will be crossed already by 
the first phase of LIGO and VIRGO.
The figure also shows that with LISA 
we will start probing a substantial
range of values of the overall coefficient of the scenario
with random-walk noise levels suppressed by the square of the Planck 
length. Values of this coefficient down to $c/(\lambda W^2)$
will be probed (in fact, even the line ``RW2'', which 
corresponds to the mentioned
scenario $\rho_h \sim c L_p^2/(\lambda W^2 f^2)$,
will be, at least marginally, probed by LISA).}
\label{fig}
\end{figure}

\end{document}